\documentclass[cits]{PoS}
\usepackage{subfigure}
\usepackage{amsmath}
\usepackage{amssymb}
\title{Nucleon electromagnetic form factors with 2+1 flavors of domain wall fermions}

\ShortTitle{Nucleon electromagnetic form factors with 2+1 flavors of DWFs}

\author{\speaker{M.~F.~Lin}$^a$\thanks{Present affiliation: Department of Physics, Yale University, New Haven, CT 06520, USA},
J.~D.~Bratt$^a$,
M.~Engelhardt$^b$,
Ph.~H\"agler$^c$,
T.~R.~Hemmert$^d$,
H.~B.~Meyer$^a$,
J.~W.~Negele$^a$,
A.~V.~Pochinsky$^a$,
M.~Procura$^a$,
W.~Schroers$^e$,
S.~Syritsyn$^a$\\
\llap{$^a$}Center\hspace{-0.0725pc} for\hspace{-0.0725pc} Theoretical\hspace{-0.0725pc} Physics,\hspace{-0.0725pc} Massachusetts\hspace{-0.0725pc} Institute\hspace{-0.0725pc} of\hspace{-0.0725pc} Technology,\hspace{-0.0725pc} Cambridge,\hspace{-0.0725pc} MA\hspace{-0.0725pc} 02139,\hspace{-0.0725pc} USA\\
\llap{$^b$} Physics Department, New Mexico State University, Las Cruces, NM 88003-8001, USA\\
\llap{$^c$} Institut f\"ur Theoretische Physik T39, Physik-Department der TU M\"unchen,
James-Franck-Stra\ss{}e, D-85747 Garching, Germany\\
\llap{$^d$} Theoretische Physik, Universit\"at Regensburg, D-93040 Regensburg, Germany \\
\llap{$^e$} Institute of Physics, Academia Sinica, Taipei 115, Taiwan
}

\abstract{
We present the recent high-statistics calculations of the nucleon electromagnetic form factors with fully dynamical domain wall fermions on the $32^3\times64$ lattices generated by the RBC and UKQCD collaborations, with pion masses at roughly 297 MeV, 355 MeV and 403 MeV. We study the phenomenological fits to the momentum transfer dependence of the form factors and investigate chiral extrapolations for the Dirac radius, Pauli radius and the anomalous magnetic moment using two variants of chiral effective field theories, the small scale expansion (SSE) and covariant baryon chiral perturbation theory. 
         }

\FullConference{The XXVII International Symposium on Lattice Field Theory - LAT2009\\
		 July 26-31 2009\\
		 Peking University, Beijing, China}

\begin{document}

\section{Introduction}
Nucleon electromagnetic form factors contain information about the size, shape and current distributions inside a nucleon. Conventionally the Dirac ($F_1$) and Pauli ($F_2$) form factors are used to parametrize the nucleon electromagnetic matrix element through the following definition:
\begin{eqnarray}
  \langle N(P') | J_{em}^\mu(x) | N(P) \rangle & = & e^{i(P'-P)\cdot x} \bar{u}(P') 
  \left [ \gamma^\mu {F_1(Q^2)} 
    + i \sigma^{\mu\nu} \frac{q_\nu}{2 M_N} {F_2(Q^2)} \right ] u(P), 
\end{eqnarray}
where $Q^2 = -q^2 = -(P'-P)^2$ is the momentum transfer of the nucleon. $J_{em}^\mu(x)$ is the electromagnetic current, with the explicit form $ J^\mu_{\mathrm{em},p} = \frac{2}{3}\bar{u}\gamma^\mu u - \frac{1}{3}\bar{d}\gamma^\mu d$ for the proton and $J^\mu_{\mathrm{em},n} = -\frac{1}{3}\bar{u}\gamma^\mu u + \frac{2}{3}\bar{d}\gamma^\mu d$ for the neutron.

Sachs electric ($G_E$) and magnetic ($G_M$) form factors are also frequently used by experimentalists, and they are defined as the following linear combinations of the Dirac and Pauli form factors:
\begin{eqnarray}
  G_E(Q^2) &=& F_1(Q^2) - \frac{Q^2}{4M_N^2} F_2(Q^2), \\
  G_M(Q^2) &=& F_1(Q^2) + F_2(Q^2).
\end{eqnarray}

The isovector ($F_{1,2}^v$) and isoscalar ($F_{1,2}^s$) form factors are defined, respectively, as
    \begin{eqnarray}
      F_{1,2}^v(Q^2) &= &F_{1,2}^p(Q^2) - F_{1,2}^n(Q^2) = 
      F_{1,2}^u(Q^2) - F_{1,2}^d(Q^2) = F_{1,2}^{u-d}(Q^2), \\
      F_{1,2}^s(Q^2) &= & F_{1,2}^p(Q^2) + F_{1,2}^n(Q^2) = 
      \frac13\left(F_{1,2}^u(Q^2) + F_{1,2}^d(Q^2)\right) = \frac13 F_{1,2}^{u+d}(Q^2),
    \end{eqnarray}
where $F_{1,2}^u$ and $F_{1,2}^d$ are Dirac and Pauli form factors which parametrize the up and down quark vector currents, respectively. Calculations of the isoscalar form factors involve the evaluation of disconnected quark loops in the three-point correlation functions and are numerically expensive. In this talk we summarize the highlights of the recent results~\cite{Syritsyn:2009mx} for the isovector form factors, in which the disconnected quark loop contributions cancel in the exact isospin limit. In particular, we discuss the phenomenological dipole and tripole fits to the $Q^2$ dependence of the form factors, and study the chiral extrapolations using the formula from the small-scale expansion (SSE) and the covariant baryon chiral perturbation theory (BChPT). Similar studies with the mixed-action approach (domain wall valence on an Asqtad sea) can be found in Refs.~\cite{Bratt:2008uf,Schroers:lat09}.

\section{Computational Details}
The calculations were performed on three $32^3\times64$ 2+1-flavor domain wall fermion gauge ensembles generated by the RBC and UKQCD collaborations~\cite{Scholz:2008uv,RBC32c} with the input light quark masses set to $am_l = 0.004, 0.006$ and $0.008$ and the strange quark mass fixed to $am_s = 0.03$. The corresponding pion masses are roughly 297 MeV, 355 MeV and 403 MeV. The Iwasaki gauge action was used with $\beta = 2.25$, which gives a lattice spacing of $a = 0.084$ fm~\cite{Syritsyn:2009mx}. The extent of the fifth dimension, $L_s$, was chosen to be 16, and the domain wall height was $M_5 = 1.8$. The choice of these parameters gives rise to a residual mass of $am_{\mathrm{res}} = 0.000660(8)$ in the chiral limit, which is about 1/6 of the lightest input quark mass. A coarse ensemble with $\beta = 2.13$ and $am_l/am_s=0.005/0.04$ on the $24^3\times64$ lattice, with a pion mass of roughly 330 MeV, was also analyzed to study the discretization errors. On this ensemble, the residual mass was determined to be $am_{\rm res} = 0.00315(1)$, and the lattice spacing was found to be $a = 0.114$ fm~\cite{Allton:2008pn}. 

We computed the forward quark propagators with a Gaussian smeared source constructed from APE smeared gauge links. The parameters for the Gaussian smearing and the APE smearing were carefully tuned to minimize the overlap with the excited states and reduce the fluctuations from the source itself (see the Appendix in Ref.~\cite{Syritsyn:2009mx} for details.). On each gauge configuration we calculated forward quark propagators at time slices $t = 10, 26, 42$ and $58$. The source-sink separation was chosen to be 12 for the fine ensembles, and 9 for the coarse ensemble, which both amount to about 1 fm in physical units. The \emph{coherent sink} technique~\cite{Bratt:2008uf} was employed to calculate the four sequential propagators associated with the four source locations simultaneously, leading to a factor of 4 reduction in the computation time. In this technique contaminations from other sinks are averaged to zero over the gauge configurations. We have verified that the results from the coherent-sink calculations were consistent with those from the independent-sink calculations, in which the sequential propagators were computed independently for each source. The form factors were obtained from the nucleon three-point functions using the overdetermined analysis as described in detail in Ref.~\cite{Syritsyn:2009mx}. And we obtained the vector current renormalization constant $Z_V$ by setting $Z_V F_1^v(0) = 1$ for each ensemble.

\section{Phenomenological Fits to the $Q^2$ Dependence}
\begin{figure}[htbp]
\centering
  \includegraphics[width=\textwidth,clip]{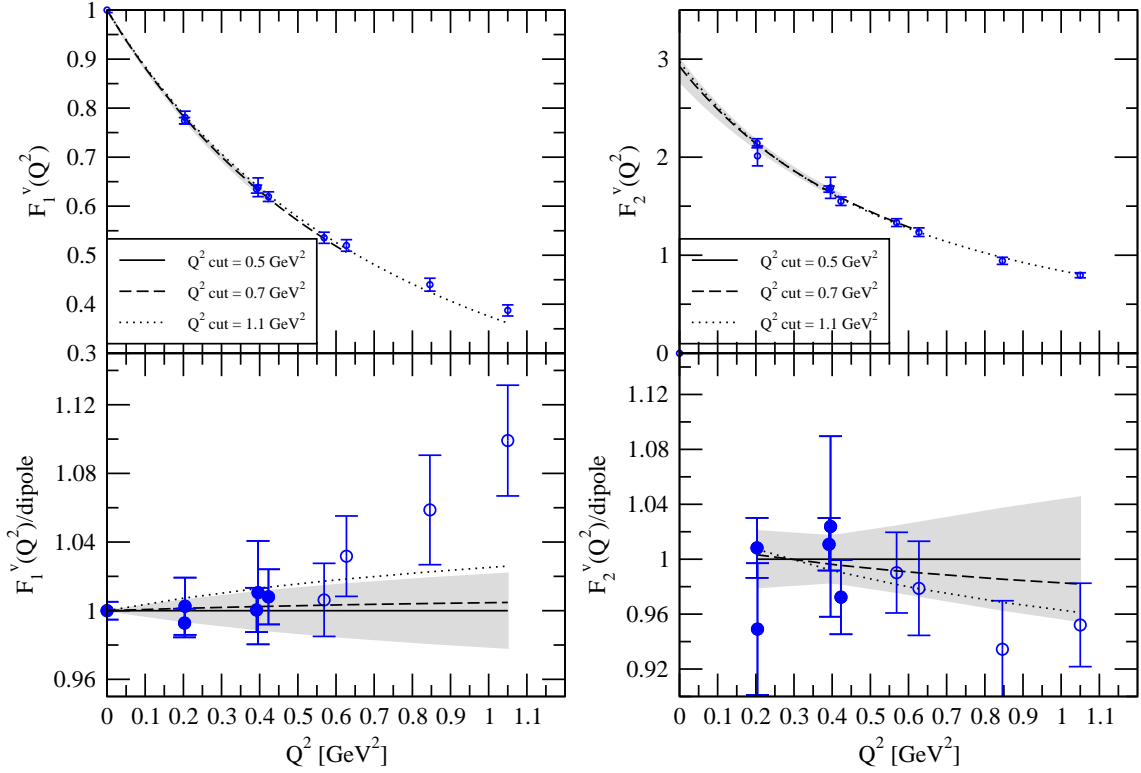}
  \label{fig:F1vF2vdipole}
\caption{Dipole fits to $F_1^{v}(Q^2)$ and $F_2^v(Q^2)$ with three different cutoffs in $Q^2$, 0.5, 0.7 and 1.1 GeV$^2$, on the $m_\pi = 297$ MeV ensemble. The top two graphs show the actual dipole fits to the lattice data. The bottom two graphs show the ratios of the data to the dipole fits with $Q^2 \leq 0.5$ GeV$^2$. Fits with the other two fit ranges are normalized relative to the $Q^2 \leq 0.5$ GeV$^2$ fits.}
\end{figure}

The $Q^2$ dependence of the isovector electromagnetic form factors is often found to be well described by a dipole form over a large $Q^2$ range. Since there is no theoretical foundation for such a $Q^2$ dependence, we performed fits to the $Q^2$ dependence of $F_1^v$ to the one-parameter dipole or tripole form of
\begin{equation}
F_1^v(Q^2) = \frac{1}{(1+Q^2/M_1^2)^n},\,\,\, n=2, 3, 
\end{equation}
while $F_2^v$ was fit to the two-parameter dipole or tripole form of
\begin{equation}
F_2^v(Q^2) = \frac{F_2^v(0)}{(1+Q^2/M_2^2)^n},\,\,\, n=2, 3. 
\end{equation}
We saw no statistically significant difference between the dipole and tripole fits. We also investigated stability of the fits by varying the maximum $Q^2$ values included in the fits, as shown in Fig~\ref{fig:F1vF2vdipole}. While the fit quality decreases with larger $Q^2$ cutoffs, the fit parameters do not show statistically significant changes. Another feature of the dipole fits is that at small $Q^2$ values, the Dirac form factor appears to be lower than the dipole fit, while at large $Q^2$, it tends to be higher, which is consistent with the observation of the experimental data~\cite{Friedrich:2003iz,Arrington:2007ux}. The Dirac and Pauli mean-squared radii, $(r_1^v)^2$ and $(r_2^v)^2$, and the anomalous magnetic moment $\kappa_v$, were obtained from the dipole fits with a $Q^2$ cutoff of 0.5 GeV$^2$ through
\begin{eqnarray}
(r_i^v)^2 = \frac{12}{M_i}, i = 1, 2; \,\,\, \kappa_v = F_2^v(0).
\end{eqnarray}
Similar behaviors of the $Q^2$ dependence were observed for the Sachs form factors $G_E^v(Q^2)$ and $G_M^v(Q^2)$. We show the results for $G_E^v(Q^2)$ from all the four ensembles in Fig.~\ref{fig:Geumd} where the phenomenological parametrization~\cite{Kelly:2004hm} of the experimental results is also plotted for comparison. 

\section{Chiral Extrapolations}
\begin{figure}[htbp]
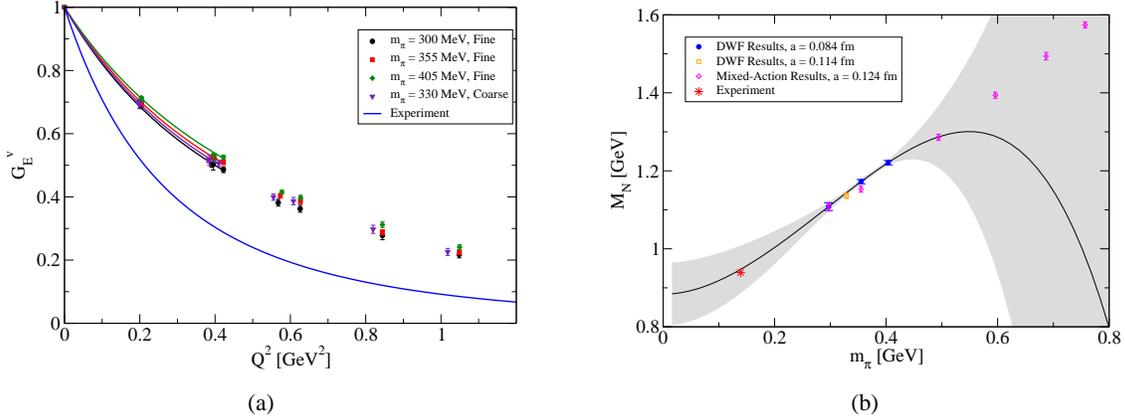

\centering
\subfigure[]{
  \includegraphics[width=0.45\textwidth,clip]{figs/Geumd_all_comp_w_exp.eps}
  \label{fig:Geumd}
}
\hfill
\subfigure[]{
  \includegraphics[width=0.45\textwidth,clip]{figs/MN_cov_extrap.eps}
  \label{fig:MN}
}
\caption{(a) Comparison of lattice results for the isovector Sachs electric form factor $G_E^v$ with the phenomenological parametrization of the experimental results~\cite{Kelly:2004hm}. (b) Chiral extrapolations of the nucleon mass using $\mathcal{O}(p^4)$ SU(2) covariant baryon chiral perturbation theory. Only the $a = 0.084$ fm domain wall results (solid circles) are included in the fits.}
\end{figure}

To compare the lattice results for the form factors at non-zero momentum transfer, we need to do chiral extrapolations at finite $Q^2$ values. Chiral perturbation theory requires that the momentum transfer values in the chiral expansions are small compared to the chiral scale of about 1 GeV. The available non-zero $Q^2$ values in our simulations range from 0.2 to 1.05 GeV$^2$, which makes it unreliable to utilize the chiral formula for the $Q^2$ dependence of the form factors. Instead, we performed chiral fits to $(r_1^v)^2$, $(r_2^v)^2\cdot \kappa_v$ (to get rid of the explicit $\kappa_v$ dependence in the chiral formula) and $\kappa_v$. As discussed in Refs.~\cite{Syritsyn:2009mx, Gockeler:2003ay}, the values of $\kappa_v$ used in the chiral extrapolations should be rescaled with a factor of $M_N^{lat}/M_N^{phys}$ to get rid of the pion mass dependent magneton. We used two variants of the chiral effective field theories to perform the extrapolations. One is the heavy baryon chiral perturbation theory with the explicit $\Delta$ degrees of freedom~\cite{Bernard:1998gv}, the so-called small-scale expansion (SSE), to $\mathcal{O}(\epsilon^3)$. The other is the $\mathcal{O}(p^4)$ SU(2) covariant baryon chiral perturbation theory (BChPT)~\cite{Hemmert:lat09,Gail:2007phd}. In order to disentangle the investigations of the applicability of the chiral effective field theories from the discretization effects, we included only the fine domain wall results in the chiral extrapolations discussed below. 

\begin{figure}[htbp]
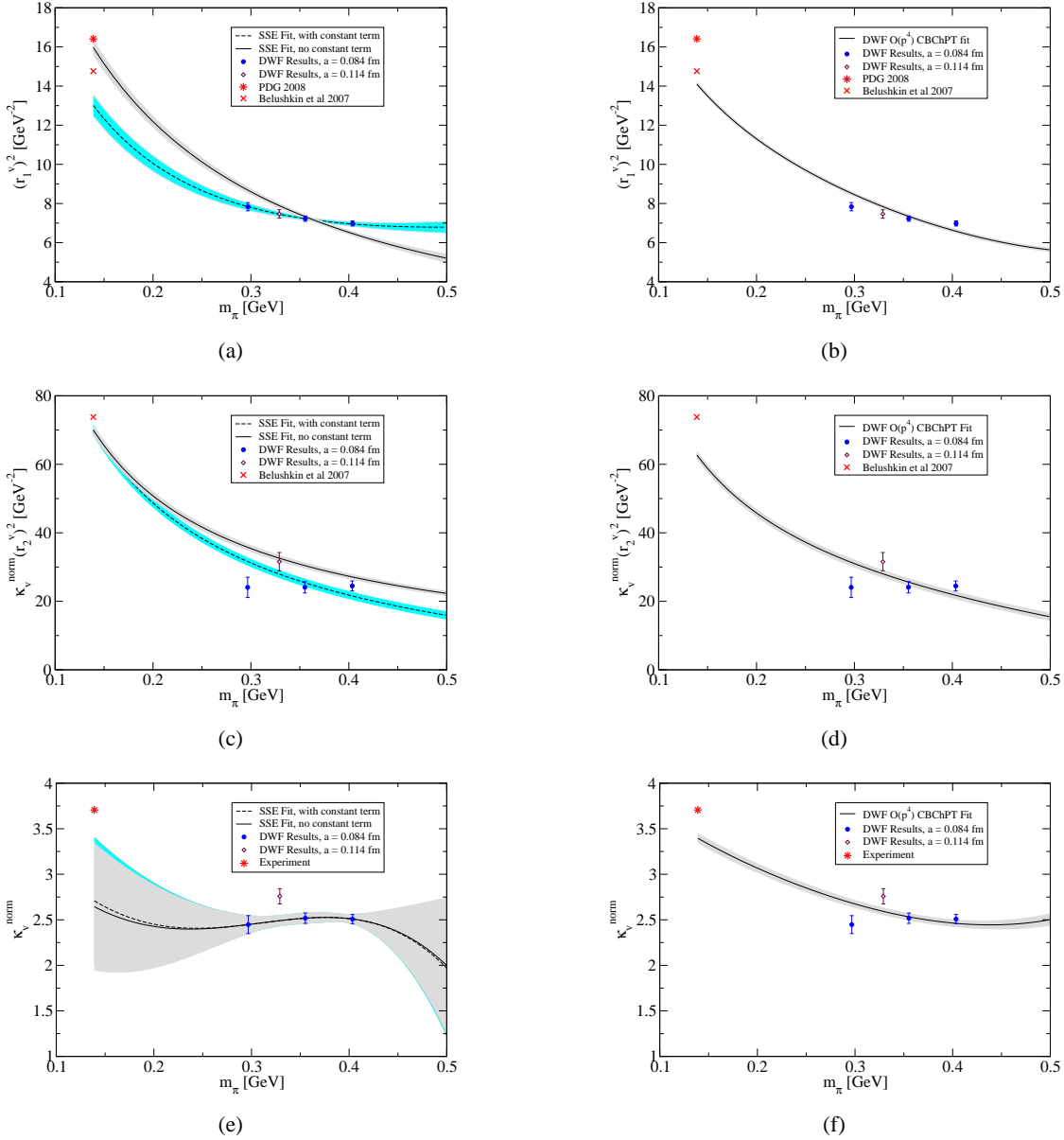

\centering
\subfigure[]{
  \includegraphics[width=0.42\textwidth,clip]{figs/r1v_SSE.eps}
  \label{fig:r1v-SSE}
}
\hfill
\subfigure[]{
  \includegraphics[width=0.42\textwidth,clip]{figs/r1v_cov.eps}
  \label{fig:r1v-cov}
}\\
\subfigure[]{
  \includegraphics[width=0.42\textwidth,clip]{figs/r2v_SSE.eps}
  \label{fig:r2v-SSE}
}
\hfill
\subfigure[]{
  \includegraphics[width=0.42\textwidth,clip]{figs/r2v_cov.eps}
  \label{fig:r2v-cov}
}\\
\subfigure[]{
  \includegraphics[width=0.42\textwidth,clip]{figs/kappa_SSE.eps}
  \label{fig:kappa-SSE}
}
\hfill
\subfigure[]{
  \includegraphics[width=0.42\textwidth,clip]{figs/kappa_cov.eps}
  \label{fig:kappa-cov}
}\\
\caption{\label{fig:fits}Chiral fits for the isovector Dirac and Pauli mean-squared radii and the anomalous magnetic moment of nucleon. The left panel shows the fits using the $\mathcal{O}(\epsilon^3)$ SSE formula. The right panel shows the fits using the $\mathcal{O}(p^4)$ BChPT formula. Only the $a = 0.084$ fm results are included in the fits.}
\end{figure}

In the SSE chiral extrapolations, we performed simultaneous fits to $(r_1^v)^2$ and $(r_2^v)^2\cdot \kappa_v$ (solid lines in Figs.~\ref{fig:r1v-SSE} and \ref{fig:r2v-SSE}), leaving the pion-nucleon-$\Delta$ coupling constant, $c_A$, and a counter term, $B_{10}^r(\lambda)$, as free parameters, while other low-energy constants were fixed to their phenomenological values. Then the value of $c_A$ obtained from such fits was used as an input to determine three other unknown parameters ($c_V, \kappa_v^0, E_1^r(\lambda)$) in $\kappa_v$ (solid line in Fig.~\ref{fig:kappa-SSE}).  As seen from the figures, the fits do not describe the lattice data very well. Adding a constant term to the formula for $(r_2^v)^2\cdot \kappa_v$ (dashed lines in Figs.~\ref{fig:r1v-SSE} and \ref{fig:r2v-SSE}) improves the fit quality, but the resulting extrapolated values at the physical point miss the empirical results~\cite{Amsler:2008uo, Belushkin:2007dn} by 10-20\%. The difficulty is that the lattice data show a weaker curvature than the SSE expansion at the given order,  which can be a result of i) the pion masses are still too heavy for the SSE formula to be accurate to the 10-20\% level at the given order; or ii) the curvature of the data is obscured by un-controlled systematic errors, such as finite volume effects.

 A prerequisite to use the BChPT formula is to know the pion mass dependence of the nucleon mass, $M_N(m_\pi)$. We determined some of the low-energy constants needed in the  $\mathcal{O}(p^4)$ BChPT formula for $M_N(m_\pi)$ from our three fine domain wall points, as shown in Fig.~\ref{fig:MN}. In the BChPT chiral extrapolations, we performed simultaneous fits to $(r_1^v)^2$, $(r_2^v)^2\cdot \kappa_v$ and $\kappa_v$ with four free parameters. The resulting fits are shown in Figs.~\ref{fig:r1v-cov}, \ref{fig:r2v-cov} and \ref{fig:kappa-cov}. Once again, the data show less curvature than the BChPT expansion at this given order. And the extrapolated physical values are also 10-20\% lower than the empirical results.

\section{Conclusions}
We calculated the nucleon electromagnetic form factors with 2+1 flavors of domain wall fermions on three fine ensembles with $a = 0.084$ fm, and one coarse ensemble with $a = 0.114$ fm. Focuses have been given to the study of the phenomenological dipole fits to the momentum transfer dependence of the isovector Dirac and Pauli form factors, and the investigations of chiral extrapolations to the isovector Dirac and Pauli mean-squared radii and the anomalous magnetic moment of the nucleon. We used two formulations of the baryon chiral effective field theories to describe our data, the SSE formulation and the covariant baryon chiral perturbation theory, and found that neither of these formulations can describe our data well. This may be caused by the relatively heavy pion masses in our simulations or the uncontrolled systematic errors, such as the finite volume effects at the lightest pion mass. To address these questions, we may need to simulate at several lighter pion masses at several different volumes to pin down the systematic errors. This remains a task for future investigations. 

\section*{Acknowledgments}
This work is supported in part by the U.~S.~Department of Energy under Grants DE-FG02-94ER40818,
DE-FG02-05ER25681, and DE-FG02-96ER40965.
Ph.~H.~acknowledges support by the Emmy-Noether program and the cluster of excellence
``Origin and Structure of the Universe'' of the DFG,
M.~P.~acknowledges support by a Feodor Lynen Fellowship from the Alexander von Humboldt Foundation,
and T.~R.~H.~is supported by DFG via SFB/TR 55.
W.~S.~wishes to thank the Institute of Physics at Academia Sinica for their kind hospitality and
support as well as  Jiunn-Wei~Chen at National Taiwan University and Hsiang-nan~Li
at Academia Sinica for their hospitality and for valuable physics discussions and suggestions. M.~F.~L. thanks the support of Department of Physics at Yale University where the manuscript is written.
Computations for this work were carried out using the Argonne Leadership Computing Facility at
Argonne National Laboratory, which is supported by the Office of Science
of the U.~S.~Department of Energy under contract DE-AC02-06CH11357;
using facilities of the USQCD Collaboration, which are funded by the Office of Science
of the U.~S.~Department of Energy;
and using resources provided by the New Mexico Computing Applications Center (NMCAC) on Encanto.
The authors also wish to acknowledge use of  dynamical domain wall configurations
and universal propagators calculated by the RBC and LHPC collaborations and the use of
Chroma SciDAC software.

\bibliography{paper}

\bibliographystyle{JHEP-2}

\end{document}